\documentclass[aps,prl,superscriptaddress,twocolumn,floatfix,nofootinbib,preprintnumbers,showpacs]{revtex4}
\usepackage{amsmath,multirow}
\usepackage{graphicx,tabularx,epsfig}

\newcommand{\etal}{\textit{et al.}}
\newcommand{\ie}{{\sl i.e. }}
\newcommand{\eg}{{\sl e.g. }}

\newcommand{\gev}{{\ensuremath\rm GeV}}

\begin{document}

\preprint{Edinburgh 2008/16}
\preprint{PSI-PR-08-05}


\title{Supersymmetric Higgs Bosons in Weak Boson Fusion}

\author{Wolfgang Hollik}
\affiliation{Max-Planck-Institut f\"ur Physik (Werner-Heisenberg-Institut),
            Munich, Germany}

\author{Tilman Plehn}
\affiliation{SUPA, School of Physics, University of Edinburgh, 
             Scotland}

\author{Michael Rauch}
\affiliation{SUPA, School of Physics, University of Edinburgh, 
             Scotland}

\author{Heidi Rzehak}
\affiliation{Paul Scherrer Institut, Villigen PSI, Switzerland}

\begin{abstract}
  We compute the complete supersymmetric next-to-leading order
  corrections to the production of a light Higgs boson in weak boson
  fusion. The size of the electroweak corrections 
  is of similar order
  as the next-to-leading order corrections in the Standard Model. The
  supersymmetric QCD corrections turn out to be significantly smaller
  than their electroweak counterparts. These higher--order corrections
  are an important ingredient to a precision analysis of the
  (supersymmetric) Higgs sector at the LHC, either as a known
  correction factor or as a contribution to the theory error.
\end{abstract}

\maketitle


The main task of the LHC era is to understand electroweak
symmetry breaking and the ultraviolet completion of the Standard
Model. According to electroweak precision data we expect to see a
light Higgs boson, which should be embedded into a UV completion
solving the hierarchy problem. A minimal realization of TeV-scale
supersymmetry (MSSM) is a leading candidate for that.
The most promising discovery channel for a light supersymmetric Higgs
boson is the production in weak-boson fusion with a subsequent decay
into tau leptons~\cite{wbf_ex,wbf_tau}.  There, a light
supersymmetric Higgs scalar is guaranteed to appear over the entire
MSSM parameter space~\cite{wbf_susy}.

In the Standard Model as well as in the MSSM, many years of LHC
running will be devoted to understanding the Higgs sector in detail,
for example extracting the gauge and Yukawa couplings~\cite{duhrssen}.
To meaningfully distinguish between, for example, the Standard-Model and
the MSSM Higgs sectors~\cite{abdel} we need to control the theory
error on the LHC rates including higher-order effects.
In the Standard Model the next-to-leading order QCD and electroweak
corrections to weak-boson-fusion Higgs production are known to be
fairly small~\cite{wbf_nlo}. In particular, the QCD corrections are
suppressed due to the color structure of the
production process and the forward-jet topology.
Also interference effects between Higgs production in weak-boson fusion and
in gluon fusion with two additional jets are strongly 
suppressed~\cite{gVBF_interf}.

For either a comparison between the Standard-Model and the MSSM Higgs sectors
or for a precision analysis of the MSSM Higgs sector these
higher-order corrections have to be augmented by supersymmetric particle
loops. In parallel to the Higgs searches, the LHC experiments will
also search for direct signatures of new physics. If we
should find such new states we can then predict their effects on the
Higgs sector. If, for example, squarks and gluinos should be
too heavy or the spectrum should be not favorable to precision
MSSM analyses~\cite{sfitter}, we need to include their effects in the
theory errors. Both cases require a comprehensive
calculation of the supersymmetric contributions to the weak-boson-fusion
and gluon-fusion production processes~\cite{maggi}.\bigskip

\underline{Supersymmetry vs Standard Model} --- 
Compared to its Standard-Model counter part the leading-order
production rate of a light supersymmetric Higgs scalar $h^0$ includes
an additional coupling factor $\sin(\beta-\alpha)$. It is expressed in
terms of the ratio of the vacuum expectation values $\tan \beta$ and
the scalar mixing angle $\alpha$ from the supersymmetric
two-Higgs-doublet model. For a given Higgs mass we can relate the
tree-level MSSM production rate to the Standard-Model result via a
simple re-scaling by $\sin^2(\beta-\alpha)$. For 
large pseudoscalar Higgs masses $m_A \gtrsim 200~\gev$ this factor is
very close to unity.

Including higher orders, there are additional contributions from the
supersymmetric particle spectrum: first, we take into account loops of
supersymmetric partners. If we assume $R$ parity, one-loop diagrams
cannot mix supersymmetric and Standard-Model particles, allowing for
a diagram-by-diagram separation of the MSSM contributions.
Secondly, the additional supersymmetric Higgs bosons with their
Standard-Model type $R$ charge appear in loops. Because the
Standard-Model Higgs boson does not simply correspond to one
supersymmetric Higgs boson we cannot separate the
Standard-Model Feynman diagrams from the MSSM set. Instead,
we first compute the MSSM-Higgs corrections and then
subtract the Standard-Model Higgs loops, scaled by the tree-level
correction factor.

Because of the large number of Feynman diagrams we compute the cross
sections using the automated tool HadCalc~\cite{hadcalc}. The Feynman
diagrams and the amplitudes we generate with
FeynArts/FormCalc~\cite{feynarts}.  The loop integrals we evaluate using
LoopTools~\cite{looptools}.  We assume minimal flavor violation, because
after taking into account all experimental and
theoretical constraints, the effect of non-minimal flavor violation on
LHC rates is small~\cite{lhc_nmfv}. We also assume a
$CP$-conserving MSSM.\bigskip

\begin{table}[t]
\begin{tabular}{l|rr|rr}
& \multicolumn{2}{c|}{effective theory}  
& \multicolumn{2}{c}{Feynman diagrams} \\
& $\alpha_{\text{eff}}$ & 
  full                  & 
  $\alpha_{\text{eff}}$ & 
  full                   \\
\hline
$\lambda_{HHH}$ 
& 
$0.208$ & 
$0.198$ & 
$0.210$ & 
$0.210$ \\
$\lambda_{HHh}$ 
& 
$-0.285$ & 
$-0.275$ & 
$-0.284$ & 
$-0.279$ \\
$\lambda_{Hhh}$ 
& 
$-0.216$ & 
$-0.219$ & 
$-0.220$ & 
$-0.257$ \\
$\lambda_{hhh}$ 
& 
$0.952$ &
$1.503$ &
$0.950$ &
$1.276$ \\
\hline
$\alpha_{\text{eff}}$ 
& \multicolumn{2}{c|}{$-0.1132$} 
& \multicolumn{2}{c }{$-0.1158$} \\
$m_h$ 
& \multicolumn{2}{c|}{$109.8~\gev$} 
& \multicolumn{2}{c }{$111.0~\gev$} \\
$m_H$ 
& \multicolumn{2}{c|}{$391.5~\gev$} 
& \multicolumn{2}{c }{$391.6~\gev$} \\
\end{tabular}
\caption{Higgs self couplings for the parameter point SPS1a 
 following Ref.~\cite{subh} (left) and 
 Ref.~\cite{feynhiggs} (right). The common factor $-3
  e m_W/(2 c_W^2 s_W)$ is not included.}
\label{tab:coupl}
\end{table}

\underline{Higgs-sector corrections} ---
The mass of the light supersymmetric Higgs boson $m_h$ is not a free
parameter. At tree level it can be computed from $m_A$ and $\tan\beta$
and is always smaller than $m_Z$. 
Higher-order corrections, dominated by
the top Yukawa coupling term $h_t^4$ at one-loop order,
push $m_h$ to values beyond the LEP2
limits~\cite{mh_feyn,feynhiggs,mh_eff,subh,mh_rec}.  Phenomenologically relevant
studies therefore need to include quantum corrections to the Higgs
mass and the Higgs potential.

The challenge in including these higher-order corrections in our
calculation is that we cannot simply shift the final-state Higgs
mass. Already in the Standard Model the physical Higgs mass $m_h$ is
linked to the running quartic coupling $\lambda (Q)$ and top
Yukawa $h_t (Q)$~\cite{sirlin}
\begin{alignat}{5}
\frac{m_h^2}{\lambda(Q) v^2} =
  1 
&- \frac{3 h_t^2(Q)}{16\pi^2} 
  \left[ Z\left( \frac{4m_t^2}{m_h^2} -1
          \right) - 2 \right]
  \left[ 1 - \frac{4m_t^2}{m_h^2} \right] \nonumber \\
&- \frac{3 h_t^2(Q)}{16 \pi^2} 
  \left[ 2 - \frac{4m_t^2}{m_h^2} \right]
  \log \frac{Q^2}{m_t^2} 
\end{alignat}
with $Z(x) = 2 \sqrt{x} \tan^{-1} (x^{-1/2})$ for $x>1/4$. When
including self--energy corrections to the Higgs mass, we should also
correct the Higgs self couplings at the same order in perturbation
theory. While this has to be done explicitely~\cite{lambda_hhh}
in a Feynman-diagrammatic
approach~\cite{mh_feyn,feynhiggs} it is automatically taken care of if
we compute the quantum effects in the scalar potential~\cite{mh_eff,subh}.

On the other hand, quantum corrections to the Higgs potential are
usually computed for vanishing external momentum instead of
$p^2 = m_h^2$. In the Feynman-diagrammatic approach it is straight
forward to compute the Higgs self-energy diagrams with a finite
momentum flow, while this is a challenge for the effective-potential
method. 

Last but not least, when computing the effective scalar
potential using renormalization-group techniques, it becomes
increasingly tedious to separate scales, like the heavy Higgs mass
$m_A$, the light stop mass $m_{\tilde t_1}$ and the gluino mass
$m_{\tilde {g}}$.\smallskip

Because the light supersymmetric Higgs boson is the lightest particle
in the supersymmetric loops we expect the numerical effects of its
mass and of the associated self coupling $\lambda_{hhh}$ to be
non-negligible. In Tab.~\ref{tab:coupl} we compare the values for the
scalar self couplings in the effective theory approach (implemented
from Ref.~\cite{subh}) with those from the Feynman-diagrammatic
FeynHiggs package~\cite{feynhiggs}. Both approaches allow for an
approximate computation introducing an effective mixing angle
$\alpha_{\text{eff}}$ from the scalar Higgs mass matrix.  
In this case the trilinear Higgs coupling
$\lambda_{hhh}$ is given by the Standard-Model coupling
times the MSSM correction factor $\cos 2 \alpha_{\rm eff} \sin (\beta+\alpha_{\rm eff})$.
The results in this approximation should
be equivalent in both schemes, 
because finite values of $p^2$ as well as
corrections to the self couplings are skipped.
In Tab.~\ref{tab:coupl} we see that the $\alpha_{\text{eff}}$
approximations indeed agree very well. Between the full results, where
FeynHiggs includes the fixed-order one-loop $\mathcal{O}(h_t^4)$ 
corrections to $\lambda_{hhh}$~\cite{lambda_hhh}, we see the expected
small deviations.\smallskip

\begin{table}[t]
\begin{tabular}{l|c|c}
& 
$\Delta \sigma/\sigma(ud \to udh)$ & 
$(\sigma_{\alpha_{\text{eff}}} - \sigma_{\text{full}})/\sigma$\\
\hline
\multicolumn{3}{c}{effective theory} \\
\hline
$\alpha_{\text{eff}}$                           & $-0.389~\%$ & 
\multirow{2}{*}{$-0.122~\%$}\\
full                                            & $-0.266~\%$ & \\
\hline
\multicolumn{3}{c}{Feynman diagrams} \\
\hline
$\alpha_{\text{eff}}$                           & $-0.393~\%$ & 
\multirow{2}{*}{$-0.076~\%$}\\
full                                            & $-0.317~\%$ & \\
\hline
\multicolumn{3}{c}{Feynman diagrams, loop-improved $Z_{\text{FH}}$} \\
\hline
$\alpha_{\text{eff}}$                           & $-0.343~\%$ & 
\multirow{2}{*}{$-0.115~\%$}\\
full                                            & $-0.228~\%$ & \\
\end{tabular}
\caption{Schemes for computing the supersymmetric corrections
  to the $VVh$ vertices at the hadronic level, for the
  leading partonic subprocess.}
\label{tab:higgs}
\end{table}

For LHC cross sections this comparison is complicated by the different
final-state Higgs masses in the two schemes.  The effect of all
supersymmetric corrections to the $VVh$ vertices ($V=W,Z$) and the
Higgs wave-function renormalization for the dominant partonic
subprocess $ud \to udh$ we show in Tab.~\ref{tab:higgs}.  The
contributions from the Higgs sector and from supersymmetric particles
cannot be separated because of their combined renormalization.
The first four lines use a
wave-function renormalization at one loop.
In the $\alpha_{\text{eff}}$ approximation the
corrections agree well, despite the different external Higgs masses.
For the
bottom lines we include the higher-order improved wave-function
renormalization factors provided by FeynHiggs. 

For the numerical analysis in this 
letter we use the Feynman-diagrammatic approach with one-loop $Z$ factors,
to allow for a proper renormalization-scale behavior.
The difference of $\sim 0.05~\%$ between the different rate predictions is
a lower limit on the remaining theory error from the MSSM Higgs sector.
Note,
however, that this theory error only applies if we strictly assume the
minimal renormalizable supersymmetric Higgs sector.\bigskip

\begin{figure*}[t]
\includegraphics[width=0.18\textwidth]{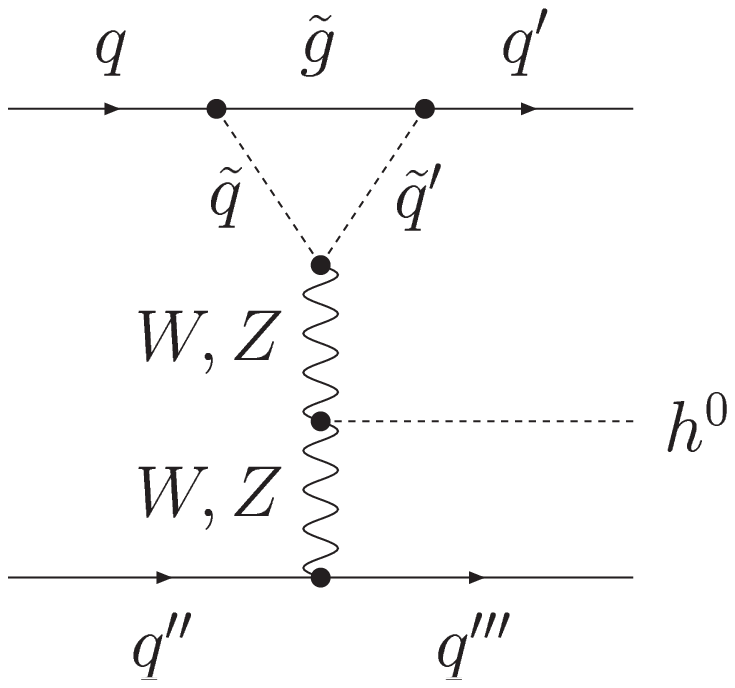}
\hspace*{5mm}
\includegraphics[width=0.20\textwidth]{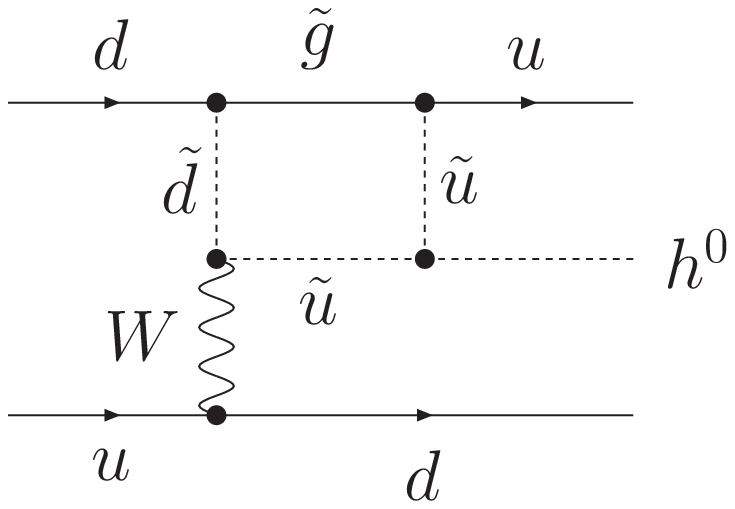}
\hspace*{5mm}
\includegraphics[width=0.20\textwidth]{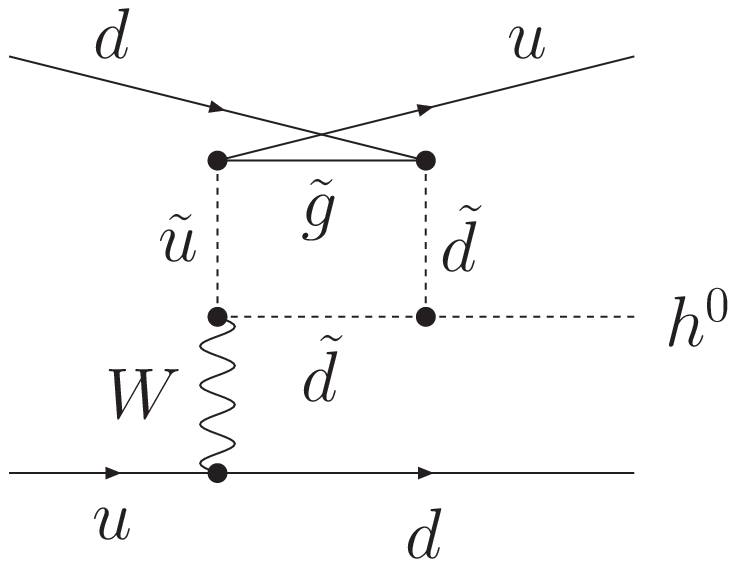}
\hspace*{5mm}
\includegraphics[width=0.22\textwidth]{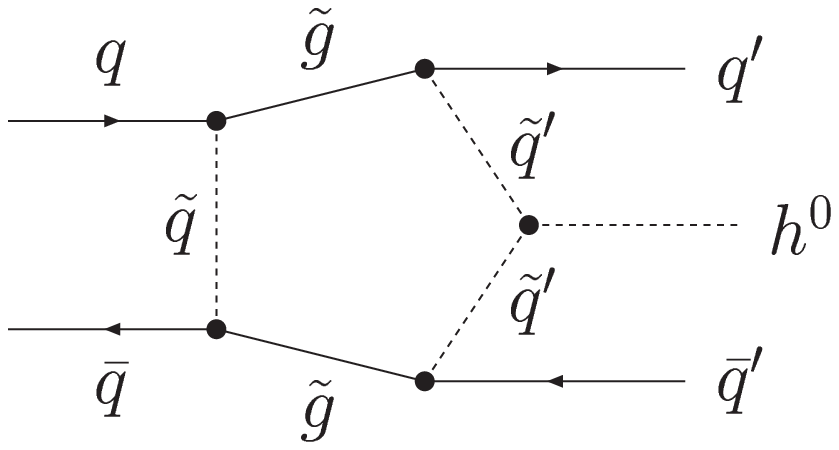}
\caption{Left to right: Feynman diagrams contributing
  to strong vertex corrections, strong boxes, strong
  pentagons.}
\label{fig:feyn}
\end{figure*}

\underline{Supersymmetric particle corrections} ---
Example one-loop diagrams appearing in the
$qq\rightarrow qqh$ processes we show in
Fig.~\ref{fig:feyn}. Our numerical results are based around the
parameter point SPS1a~\cite{sps}, for which there are two aspects 
we need to remember: supersymmetric particles with only electroweak
charges have typical masses of 100--200~GeV, while squarks and gluinos
range around 500--600~GeV, and $\tan \beta =10$ avoids large
non-decoupling effects from down-type fermions.\smallskip

As we can see in Tab.~\ref{tab:sps1a} all QCD corrections $\Delta
\sigma \propto \alpha^3 \alpha_s$ turn out to be surprisingly small.
In the Standard Model we know that from a QCD perspective we are
essentially looking at two-sided non-interfering deep inelastic
scattering.  However, there are several mechanisms responsible for an
even larger suppression in the supersymmetric case.\smallskip

Two tree-level vertices receive one-loop corrections, $qqV$ and
$VVh$, but only the first is corrected by squark/gluino loops.
The $\tilde{q} \tilde{q}'W$ coupling
connects left-handed sfermions. Since the mixing between
left and right-handed light-flavor squarks is proportional to
the negligible quark Yukawa coupling, both external quarks are then
left-handed, just as at tree level. This means that in the one-loop
diagram (when closed with the Born diagram) the left-handed fermion trace 
cannot be connected through a gluino-mass insertion, because this would
require a chirality flip.  Instead of $m_{\tilde{g}}$, 
the typical momentum scale in the
numerator is $\sim m_h/2$, an order of magnitude
below the gluino mass in the denominator.

In the electroweak case, also the (typically lighter) charginos and
neutralinos in the loop couple to the vector boson. This means that 
we can
add a double mass insertion into the fermion line which can partly
compensate for the heavy masses in the loop denominator.  
This effect leads to a
relative enhancement of the electroweak over the QCD $qqV$ vertex
correction we observe in Tab.~\ref{tab:sps1a}.\smallskip

\begin{table}[b]
\begin{tabular}{lr||lr}
\hline
diagram & $\Delta \sigma/\sigma~[\%]$ &
diagram & $\Delta \sigma/\sigma~[\%]$ \\
\hline
\multicolumn{2}{c||}{$\Delta \sigma \sim \mathcal{O}(\alpha)$}  &
\multicolumn{2}{c  }{$\Delta \sigma \sim \mathcal{O}(\alpha_s)$} \\
\hline
self energies      &  0.199    && \\
\hline
$qqW+qqZ$          & -0.392    &
$qqW+qqZ$          & -0.0148      \\
$qqh$              & -0.0260   &
$qqh$              &  0.00545     \\
$WWh+ZZh$          & -0.329    && \\
\hline
box                &  0.0785   &
box                & -0.00518     \\
pentagon           &  0.000522 &
pentagon           & -0.000308    \\ 
\hline
\multicolumn{4}{c}{sum of all $\Delta \sigma/\sigma = -0.484~\%$}
\end{tabular}
\caption{Complete supersymmetric 
  corrections to the process $pp \to qqh$ by diagrams. 
  Our parameter point SPS1a has a 
  tree--level rate of 706~fb.}
\label{tab:sps1a}
\end{table}

For strongly interacting boxes, the 
$\tilde{q} \tilde{q}' W$ and $q \tilde{q} \tilde{g}$ couplings are
the same for both diagrams shown in Fig.~\ref{fig:feyn}, but the
$\tilde{q} \tilde{q} h$ coupling is proportional to $T_3 - Q s_w^2$, \ie
around $-1/3$ for down squarks and $+5/16$ for up squarks.
This leads to a cancellation by one order of
magnitude, which could only be broken by different squark masses.
Left-handed squarks, however, form a $SU(2)$ doublet and are governed
by the same soft-breaking terms, and the left--right mixing is
negligible for light--flavor squarks. This argument does not hold for
the sub-leading $ZZ$ fusion, where we indeed find that the corrections
turn out to be at a more natural level.\smallskip

In the Standard Model the color factor of a gluon exchange between the
two incoming quarks is proportional to the trace of the $SU(3)$
generators and hence zero. The same is true for a pentagon gluino
exchange between the incoming quarks, where the color trace is
evaluated along quark/squark lines.  In Fig.~\ref{fig:feyn}, we show
another supersymmetric pentagon diagram with a squark exchange between
the two incoming quarks. The $VV$-fusion is
replaced by a squark coupling to the Higgs, which gets rid 
of the color suppression.
Such diagrams contribute formally at
order $\mathcal{O}(\alpha_s^2 \alpha^2)$, which is as large as the
Born term $\mathcal{O}(\alpha^3)$.  However, their kinematic
properties are completely different from the
vector-boson-fusion topology and the large loop masses further
reduce their contribution to an altogether negligible level. 

Following all 
the above arguments the supersymmetric QCD corrections to
weak-boson-fusion Higgs production are suppressed by a whole list of
mechanisms, which explain their at first sight surprising suppression
even with respect to the electroweak corrections in
Tab.~\ref{tab:sps1a}.\bigskip

Looking beyond SPS1a, we show the
next-to-leading order corrections for the complete set of SPS
parameter points~\cite{sps} in Tab.~\ref{tab:sps}. From the discussion above
we do not expect the picture of electroweak vs. strong corrections
to change significantly for any of them.
Heavier supersymmetric spectra and 
different values of $\tan\beta$ and of the trilinear couplings
just scale the over-all size of the supersymmetric corrections. The 
relatively large corrections for the SPS5 parameter point are driven by a
light top squark, while the largely 
decoupled spectrum in SPS9 leads to negligible MSSM effects. The typical
size of the complete MSSM corrections is less or around 1~\%.\smallskip

\begin{table}[t]
\begin{tabular}{l||r|r|r||r}
\multicolumn{5}{c}{$\Delta \sigma/\sigma~[\%] $} \\
\hline
& $WWh+ZZh$
& $\mathcal{O}(\alpha)$
& $\mathcal{O}(\alpha_s)$
& all \\
\hline
SPS1a & -0.329 & -0.469 & -0.015 & -0.484 \\
SPS1b & -0.162 & -0.229 & -0.006 & -0.235 \\
SPS2  & -0.147 &  0.129 & -0.002 & -0.131 \\
SPS3  & -0.146 & -0.216 & -0.006 & -0.222 \\
SPS4  & -0.258 & -0.355 & -0.008 & -0.363 \\
SPS5  & -0.606 & -0.912 & -0.010 & -0.922 \\
SPS6  & -0.226 & -0.309 & -0.010 & -0.319 \\ 
SPS7  & -0.206 & -0.317 & -0.006 & -0.323 \\
SPS8  & -0.157 & -0.206 & -0.004 & -0.210 \\
SPS9  & -0.094 & -0.071 & -0.003 & -0.074 
\end{tabular}
\caption{Complete MSSM corrections for all SPS 
  parameter points~\cite{sps}. The vertex correction in the 
  first column corresponds 
  to Tab.\ref{tab:higgs}, but including all partonic channels.}
\label{tab:sps}
\end{table}

\begin{figure}[t]
\hspace*{-2ex}
\includegraphics[width=0.50\textwidth]{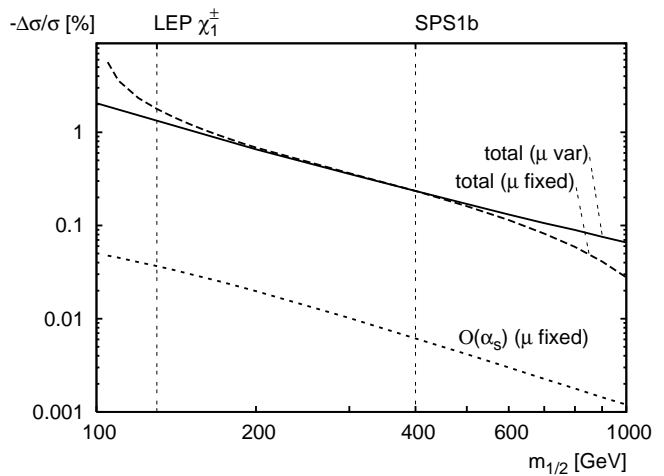}
\caption{Relative next-to-leading order corrections as a function of 
  $m_{1/2}$ for varying and for fixed $\mu$. For the latter we show
  the strong corrections independently. The vertical lines indicate
  the chargino mass limit from LEP2 and the reference point SPS1b.}
\label{fig:scan1b}
\end{figure}

To study the behavior of the one--loop corrections with varying
supersymmetric masses we start from the parameter point SPS1b and run
the universal gaugino mass $m_{1/2}$ from 100 to 1000~GeV.  In
Fig.~\ref{fig:scan1b} we show the result for a $m_{1/2}$-dependent
Higgsino mass
parameter as well as for the fixed SPS1b value $\mu = 499~\gev$. 
The corrections
sharply drop with increasing $m_{1/2}$, as we approach the decoupling
limit. Fixing $\mu$ means larger corrections for a light SUSY spectrum
and a sharper drop for heavy masses. The maximum size for the
corrections consistent with the LEP2 chargino limit we read off to be
$-2~\%$. If we tune all weak-scale MSSM parameters to barely respect
all LEP2 and Tevatron limits we find that the size of the
supersymmetric corrections is bounded by $-4\%$. Explicit
non-decoupling effects in the bottom Yukawa only appear in this
process at the two-loop level, which means all curves in
Fig.~\ref{fig:scan1b} decouple smoothly for increasing masses.
Consistent with our previous discussion the $\mathcal{O}(\alpha_s)$
corrections are negligible over the entire parameter range.\bigskip

\underline{Outlook} ---
In the light of a possible precision analysis of the Standard-Model
and MSSM Higgs sector at the LHC we have analyzed the size of the
supersymmetric one-loop corrections to the weak-boson-fusion
production process $qq \to qqh$.

The appearance of all supersymmetric neutral Higgs bosons in the loops
required us to study the impact of different methods of
describing higher-order effects on Higgs masses and the Higgs
potential. We find that the corrections from the Higgs sector are at
the per-cent level, with a remaining uncertainty of below 0.1~\% due to these
calculational approaches --- simply reflecting unknown higher-order
corrections. 

The supersymmetric one-loop QCD corrections are not only suppressed
to a typical NNLO level, but turn out to be negligible.
This is due to a variety of effects, based on the color
structure, the supersymmetric coupling structure, or the kinematics of
the process. The complete set of electroweak loop diagrams contributes
at the per-cent level, as is expected for massive $\mathcal{O}(\alpha)$
corrections. 

In total, the supersymmetric one-loop corrections to Higgs production
via vector--boson fusion can be up to 4~\% for parameter points
allowed by direct SUSY searches and are typically at or below 1~\%.
Their sign is in general negative. This result should serve as a solid
basis for a precision analysis of the supersymmetric Higgs sector at
the LHC.\smallskip

\underline{Acknowledgments} --- We would like to thank Dominik
St\"ockinger and Thomas Binoth for valuable discussions. We would also
like to thank the PhenoGrid VO and ScotGrid for providing computer
resources.  This work was supported in part by the European Community's
Marie-Curie Research Training Network under contract
MRTN-CT-2006-035505 `Tools and Precision Calculations for Physics
Discoveries at Colliders' (HEPTOOLS).



\end{document}